\documentclass[twocolumn,showpacs,showkeys,preprintnumbers,amsmath,amssymb]{revtex4}

\usepackage{graphicx}
\usepackage{dcolumn}
\usepackage{bm}

\begin{document}

\title{Observation of opto-mechanical multistability in a high Q torsion balance oscillator}

\author{F. Mueller}
\author{S. Heugel}
\author{L. J. Wang} 
\email{lwan@optik.uni-erlangen.de}
\affiliation{Max Planck Research Group, Institute of Optics, Information and Photonics, University Erlangen-Nuremberg, D-91058 Erlangen, Germany}
\homepage{http://www.optik.uni-erlangen.de}

\date{September 18th, 2007}

\begin{abstract}
We observe the opto-mechanical multistability of a macroscopic torsion balance oscillator. The torsion oscillator forms the moving mirror of a hemi-spherical laser light cavity. When a laser beam is coupled into this cavity, the radiation pressure force of the intra-cavity beam adds to the torsion wire's restoring force, forming an opto-mechanical potential. In the absence of optical damping, up to 23 stable trapping regions were observed due to local light potential minima over a range of 4 $\mu$m oscillator displacement. Each of these trapping positions exhibits optical spring properties. Hysteresis behavior between neighboring trapping positions is also observed. We discuss the prospect of observing opto-mechanical stochastic resonance, aiming at enhancing the signal-to-noise ratio (SNR) in gravity experiments.
\end{abstract}


\maketitle

The coupling of mechanical oscillator systems to the resonant radiation pressure field of a light cavity has been studied for various systems~\cite{brag1,dorsel,meystre,brag3,cohadon,metzger,sheard,rokhsari,kleckner,corbitt1,schliesser,divirgilio,corbitt2}. 
The applications under investigation range from radiation pressure dynamics and noise reduction in gravitational wave detectors~\cite{brag1,corbitt1,divirgilio}, to cooling of micro- and nanooscillators to their quantum-mechanical ground state~\cite{kleckner,schliesser}. 
A common feature of these systems is that the mechanical oscillator forms a part of a resonant optical cavity, and radiation pressure modifies the potential which the oscillator sees, locally changing its shape~\cite{meystre}. 
In most cases, this leads to the creation of additional potential minima, which serve as trapping positions for the oscillator, and due to a different potential gradient, this change will also change its intrinsic oscillation properties~\cite{meystre,corbitt2}.
Another effect in some of the systems reported recently is the presence of a velocity dependent damping or anti-damping force~\cite{rokhsari,kleckner,corbitt2}, which leads to optical \lq\lq cooling\rq\rq~\cite{kleckner} or \lq\lq heating\rq\rq~\cite{rokhsari,schliesser,corbitt2} (instability) of the systems. 
These time dependent forces arise when the intra-cavity light storage time is comparable to the mechanical oscillation period~\cite{sheard,divirgilio,corbitt2}. 
On the one hand, a damping force is desirable when cooling of the fluctuations coupled to the device is an issue. 
However, the parameter dependent anti-damping force will lead to stability problems. 
In this case, it is necessary to apply a secondary feedback mechanism~\cite{cohadon}.

In a seminal work, Dorsel et al.~\cite{dorsel} observed optical bistability of a cavity in which one mirror is suspended on a long pendulum. 
Furthermore, multistability was predicted~\cite{meystre} for pendular systems in two or three mirror configurations.
However, due to the large restoring force associated with a pendulum, only minute displacement was obtained with the large laser powers ($>100\,mW$) used in the experiment.
Another important set of pendular experiments are the suspended Fabry-Perot cavities in gravitational wave detectors~\cite{divirgilio}, where opto-mechanical effects are observed in the presence of intra-cavity light power levels in the $Watt$ range.
Avoing the disadvantages of pendular suspensions, the coupling of nano- and microoscillators to optical cavities has been successfully demonstrated. However, the results are mostly restricted to a higher frequency regime $>1\,kHz$ up to the $MHz$ range, equalling those oscillators' mechanical eigenfrequencies.

In this letter, we report the observation of opto-mechanical multistability using a torsion balance oscillator as the moving mirror.  
Moreover, an electronic feedback system is implemented to further reduce the torsion oscillator's restoring force.
With this combination, we are able to generate an all-optical potential in which the original physical properties of the torsion oscillator become almost irrelevant. 
Near each resonance of the optical cavity, the torsion oscillator will automatically seek stable positions created by the opto-mechanical potential. 
We observe up to $23$ of these trapping states, in which the oscillator shows a change in oscillation frequency (\lq\lq optical spring effect\rq\rq)~\cite{corbitt2,sheard,divirgilio}.
Applying a slowly changing force to the oscillator, we observe \lq\lq hopping\rq\rq between neighboring trapping states, either unidirectional in a staircase-like manner, or bidirectional between two or more states, also showing hysteresis behavior.
Finally, with such a multistable system, it becomes possible to investigate another interesting effect, namely the stochastic resonance between neighboring trapping states~\cite{badzey,wiesenfeld,karapetyan}. 
Exploiting this technique, it appears feasible to further enhance the signal-to-noise ratio (SNR) in similar systems, even for gravitational wave detection~\cite{karapetyan}.

\begin{figure}                                          
\includegraphics{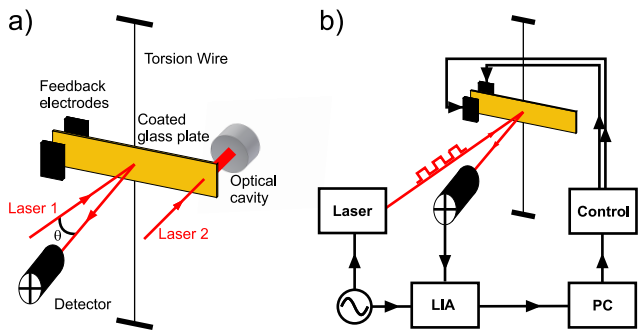}
\caption{Simplified schematic of the experiment. The torsion balance oscillator (a) is a well-known precision force measurement device, sensitive down to the fN range. In addition, there exist simple linear control techniques for the applied electrostatic feedback (b). By choosing the axis of interaction to be horizontally aligned, we avoid disturbing effects of seismic surrounding which appear in standard low-frequency pendula. Therefore, the optical coupling into the hemi-spherical cavity is stable enough to observe low order TEM cavity modes. LIA: Lock-in amplifier.}
\label{fig1}
\end{figure}

The experimental system is shown schematically in Fig.~\ref{fig1}. 
It consists mainly of a torsional oscillator~\cite{gillies} made of a gold-coated glass plate, $50\,mm\times10\,mm\times0.15\,mm$ in size, doubly suspended on a $15\,cm$ long, $25\,\mu m$ thick tungsten wire. 
The gold-coated glass plate serves as the moving flat mirror of a hemi-spherical optical cavity. 
The oscillator body has a mass of $\sim0.2\,g$ and a moment of inertia $I=4.4\,\times\,10^{-8}\,kg\,m^{2}$. 
The torsional constant is measured to be $\tau=2.2\times10^{-7}\,Nm\,rad^{-1}$. 
The torsion pendulum has a natural frequency of $f_{0}=0.36\,Hz$ with a quality factor $Q\sim2,300$.

A laser beam is reflected from the center of the oscillator and then detected by a high-sensitivity quadrant diode detector followed by a lock-in detector~\cite{lorrain}. 
The voltage signal proportional to the oscillator's angular position is digitized at a sampling rate of $5\,kHz$. 
This scheme allows for measuring the oscillator's angular position $\theta$ with an accuracy of $2\,nrad\,Hz^{-1/2}$. 
The signal is then used as the input of a computerized, digital proportional (P) and differential (D) digital control loop. 
The generated digital computer control signal is then converted to an analog output control signal. 
The end of the torsion balance opposite to the cavity side is placed between two electronic feedback electrodes. 
Since the torsion balance is electrically grounded, varying the voltages applied to the feedback electrodes gives efficient control of the balance's angular position~\cite{speake,chen3}. 
This gives the possibility of active damping (or heating) of the torsion oscillator, as well as controlling its spring constant.
In the experiment, the effective eigenfrequency can be changed easily within a range from $20\,mHz$ to $3\,Hz$. 
The apparatus is placed in a high vacuum ($10^{-7}\,mbar$) environment, giving a high quality factor $Q\sim2,300$.
The entire setup is mounted on top of an active vibration isolation system.

The hemi-spherical optical cavity is formed by the gold-coated glass surface and a second, rigidly mounted spherical mirror with a radius of curvature $25\,mm$, at a distance of $12.5\,mm$. 
Experimentally, we observe Laguerre-Gaussian TEM$_{00}$ and TEM$_{20}$ modes with a free spectral range (FSR) of the fundamental mode at $\sim13.5\,GHz$. This optical cavity has a finesse of $F=11$, giving a mirror reflectivity of $R=0.87$.

For the free torsion oscillator, the mechanical restoring force is proportional to the linear displacement, resulting in a quadratic mechanical potential. 
Now, the position-dependent radiation pressure inside an optical cavity is added. 
This cavity is approximated to be of a Fabry-Perot type. 
The position-dependent intra-cavity light power $P(x)$ produces a radiation pressure force $F(x)\propto2P(x)/c$. 
Adding the potential of this radiation pressure force for a cavity of length $d$ to that of the torsion wire, we obtain an analytic form for the total potential
\begin{widetext}
\begin{equation}
\label{pot}
U(x)=\frac{\tau}{2L^2}\,x^2-\frac{2P_0}{c(1-R^2)^2}\cdot\frac{\lambda}{2\pi\sqrt{1+\left(\frac{2F}{\pi}\right)^2}}\cdot tan^{-1}\left[\sqrt{1+\left(\frac{2F}{\pi}\right)^2}\,tan\left[\frac{2\pi(d-x)}{\lambda}\right]\right].
\end{equation}
\end{widetext}

\begin{figure}                                          
\includegraphics{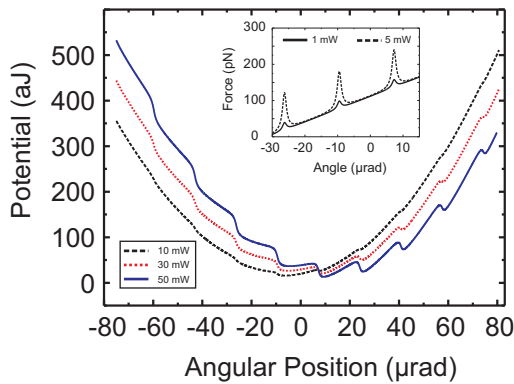}                     
\caption{Analytic calculation of the opto-mechanical potential for an amplitude range of $160\,\mu rad$. By increasing the optical input power into the cavity successively from $10\,mW$ to $50\,mW$, stable position minima are formed. This is caused by a constant torsion wire restoring force added to the repulsive cavity radiation pressure force. The inset shows the combined force on the oscillator, of which the modulated part is proportional to the intra-cavity light intensity. Here, the cavity forces are calculated for incident light powers of $1\,mW$ and $5\,mW$.}
\label{fig2}
\end{figure}

Here, $x$ is the moving mirror's linear displacement, $L$ is the balance arm's half length, $P_{0}$ is the light power incident to the cavity, $c$ is the speed of light, and $\lambda$ is the optical wavelength. 
Here, we explicitly exclude a time dependence in the opto-mechanical potential. 
This assumption is justified because the measured resonant linewidth of our cavity $\gamma\approx1.2\,GHz$ is far greater than the mechanical eigenfrequency. 
Therefore, at any given time $t$, the intra-cavity intensity is determined only by the cavity length given by the linear moving mirror displacement $x$.
Fig.~\ref{fig2} shows a plot of the opto-mechanical potential given in Eq.[~\ref{pot}] for three different optical input powers to the cavity. In this case, possible higher order TEM modes are neglected. This analytical result is very similar to that in ~\cite{dorsel}. The inset of fig.~\ref{fig2} shows the intra-cavity radiation pressure force over a smaller oscillator amplitude range, for lower light input powers of $1\,mW$ and $5\,mW$.
We find distinct force maxima at equidistant positions, if the additional cavity force field is present.
For a pure quadratic potential $U(x)$, the conservative force $F(x)$ due to the relation $F(x)=-\nabla\,U(x)$ should be linear over the oscillator's position. It is important to note that the region of maximum cavity transmission is mechanically extremely unstable. The trapping positions are located in the area of maximum force gradients.

In order to investigate the effect of multistability of the oscillator potential, we first experimentally lower its natural frequency down to $\sim70\,mHz$. 
The reason is simply that by flattening the overall mechanical potential, the potential contribution by the light field becomes dominant. 
This is achieved by applying a well-chosen negative proportional feedback signal using the digital feedback control. Effectively, this creates a \lq\lq softer spring.\rq\rq 
In addition, we supply a weak velocity-dependent damping force which cools the oscillator by removing fluctuation but does not influence the spring constant. 
This scheme is applied in all subsequent measurements.

\begin{figure}                                          
\includegraphics{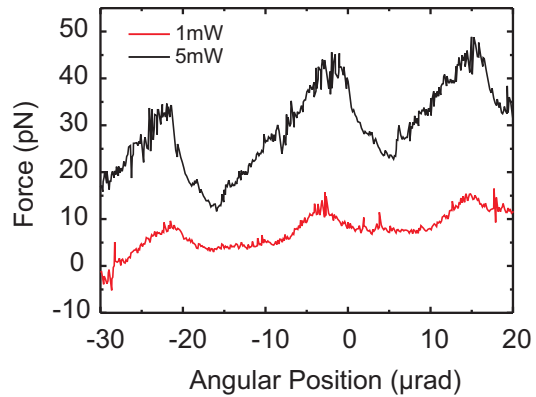}                     
\caption{The oscillator angular position is locked and then scanned over a range of $50\,\mu rad$. The simultaneous measurement of the applied electrostatic locking force reveals equally spaced maxima. The shape equals the calculation given in Eq.~\ref{pot}. The force maxima for $1\,mW$ input power appear to be symmetric, while instability prohibits the oscillator to be moved into the region of maximum cavity light power for the $5\,mW$ case. This experimental observation is in good agreement with theoretical predictions.}
\label{fig3}
\end{figure}

We first map the potential of the moving mirror by measuring the force which is necessary to hold the oscillator arm at a well-defined position. 
In the presence of the cavity light field, we use the electronic feedback loop for scanning the oscillator's angular position and simultaneously measure the equivalent electrostatic force for stabilizing the holding position. 
Fig.~\ref{fig3} shows two measurements for cavity light input powers of $1\,mW$ and $5\,mW$, respectively.
The spacing of such maxima equals the calculated and measured FSR of our cavity. In comparison to the theoretical calculation (shown in the inset of fig.~\ref{fig2}), we find good experimental agreement for the curve's shape. 
As mentioned earlier, we expect a very unstable mechanical position exactly at the cavity resonances. 
This is seen in comparison of the amplitudes of the opto-mechanical force peaks. Measured peaks are only $\sim1/3$ in amplitude compared to the calculated ones. 
In other words, in the region of maximum cavity intensity, the moving mirror does not remain stable long enough for experimental averaging.

\begin{figure}                                          
\includegraphics{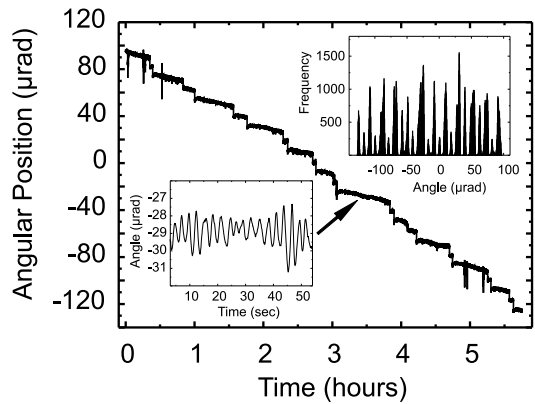}                     
\caption{The torsion wire's linear drift shows 23 stable trapping positions which are mainly caused by the TEM$_{00}$ and TEM$_{20}$ modes of the cavity. The upper inset shows a histogram plot over the full drift range, indicating that some of the main modes are split. This fact can be explained by trapping in higher order cavity modes (with lower probability). The lower inset shows a zoomed-in time trace of the central trapping position. Here, the oscillator clearly follows an optical spring behavior. Its oscillation period has changed from the soft spring period of $15\,s$ to $2.5\,s$.}
\label{fig4}
\end{figure}

If the oscillator is allowed to drift slowly after applying a constant offset voltage, it seeks a new overall equilibrium position given by the wire potential. 
Instead of a linear drift which is expected for the free torsion oscillator, the cavity-coupled system should seek locally stable positions. 
Fig.~\ref{fig4} shows the time trace of the torsion balance's angular position in this experimental configuration. 
We observe \lq\lq hopping\rq\rq of the moving mirror between neighboring trapping positions. 
Plotting a position histogram (upper inset of fig.~\ref{fig4}), we find as many as $23$ locally stable positions over a range of $\sim220\,\mu rad$ in angular position, or equivalent to $\sim4\,\mu m$ of linear displacement. 
The trapping potential minima are mainly formed by TEM$_{00}$ and TEM$_{20}$ cavity modes, equally spaced in position.
The local curvature of the opto-mechanical potential is expected to be steeper inside a trapping region. 
This effectively changes the local spring constant of the torsion oscillator.
Such a behavior can be seen in the lower inset of fig.~\ref{fig4}. 
It shows a magnified view of the time trace for the central trapping position. 
We find that the adjusted ($\sim70\,mHz$) oscillation frequency in the absence of light force now changes to a local oscillation frequency of $\sim0.4\,Hz$, indicating the well-known optical spring effect~\cite{corbitt2,sheard,divirgilio}. 
We also observe the known \lq\lq confinement\rq\rq  / cooling effect in the presence of an optical spring~\cite{corbitt2}.

\begin{figure}                                          
\includegraphics{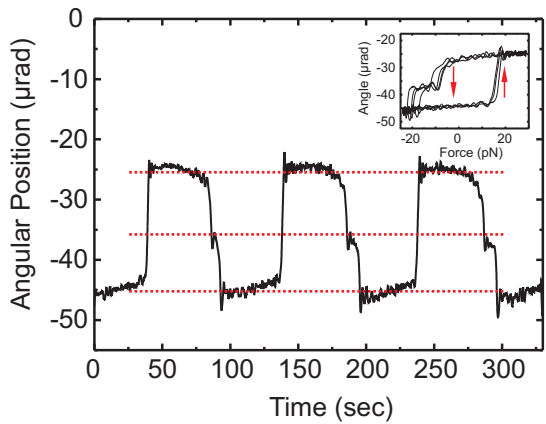}                     
\caption{The oscillator's angular position changes in discrete steps when a sinusoidal electrostatic modulation is applied. The left slope is steeper than the right one due to the asymmetric shape of the opto-mechanical potential. In addition, a higher order unstable trapping state is found on the right slope. The inset shows a plot of the angular position over the modulation force applied. The behavior of hysteresis is clearly seen.}
\label{fig5}
\end{figure}

Finally, it is interesting to investigate whether the torsion oscillator will display positional hysteresis when we allow only a limited displacement range. 
To do this, the oscillator is moved to a region containing two stable trapping positions by applying an offset voltage to the control electrodes.
In addition, a small modulation voltage of $20\,mV$ p-p in amplitude at a frequency of $10\,mHz$ is applied to the feedback electrodes.
This gives a force modulation of $52\,pN$ p-p.
The adjustment and modulation procedure induces a discrete change between two neighboring trapping positions. 
Fig.~\ref{fig5} shows the time trace of this bistable oscillator.
Instead of following the applied modulation sinusoidally, the moving mirror \lq\lq jumps\rq\rq  between two distinct positions. 
When we plot its angular position over the applied modulation force (inset of fig.~\ref{fig5}), we find a clear hysteresis in position. 
The effect of opto-mechanical position hysteresis is thus verified experimentally for this opto-mechanical system.
\newline
In conclusion, we have demonstrated a macroscopic, controlled, opto-mechanical oscillator system. 
Here, the oscillator is efficiently coupled to the resonant modes of a hemi-spherical light cavity. 
We are able to exclude the effect of velocity-dependent cooling and/or heating radiation pressure forces. 
In combination, a quasistatic multistable system is formed which exhibits a wealth of opto-mechanical effects such as optical spring effects, hopping, and position hysteresis. 
We believe that the presented experimental system is a useful test environment for further exploration of multistability effects in the low-frequency regime. 
The extent of the system's opto-mechanical coupling is linearly tunable with high stability, opening the way for additional
related experiments. 
As an example, this multistable system will be further used to investigate the effect of stochastic resonance.
Exploiting this effect, it is possible to enhance the SNR in the detection of small harmonic signals in bistable oscillators by adding noise to the system.
As was recently proposed~\cite{karapetyan}, a typical application for this technique can be found in the read-out interferometer unit of gravitational wave detectors where radiation pressure and cavity effects are non-negligible.

\begin{acknowledgments}
We thank S. Malzer and B. Menegozzi for technical support, and Z. H. Lu and T. Liu for helpful discussions.
\end{acknowledgments}

\end{document}